\documentclass{optica-article}

\journal{opticajournal} 

\articletype{Research Article}

\usepackage{lineno}
\usepackage{tabularx}
\usepackage{makecell}
\usepackage{multirow}
\usepackage{ragged2e}
\usepackage[section]{placeins}
\usepackage{float}

\begin{document}

\title{Non-volatile Programmable Photonic Integrated Circuits using Mechanically Latched MEMS: A System-Level Scheme Enabling Power-Connection-Free Operation Without Performance Compromise}

\author{Ran Tao, Jifang Qiu\authormark{*}, Zhimeng Liu, Hongxiang Guo, Yan Li and Jian Wu} 

\address{State Key Laboratory of Information Photonics and Optical Communications, School of Electronic Engineering, Beijing University of Posts and Telecommunications, Beijing 100876, China}

\email{\authormark{*}jifangqiu@bupt.edu.cn} 


\begin{abstract*} 
  Programmable photonic integrated circuits (PPICs) offer a versatile platform for implementing diverse optical functions on a generic hardware mesh. However, the scalability of PPICs faces critical power consumption barriers. Therefore, we propose a novel non-volatile PPIC architecture utilizing MEMS with mechanical latching, enabling stable passive operation without any power connection once configured. To ensure practical applicability, we present a system-level solution including both this hardware innovation and an accompanying automatic error-resilient configuration algorithm. The algorithm compensates for the lack of continuous tunability inherent in the non-volatile hardware design, thereby enabling such new operational paradigm without compromising performance, and also ensuring robustness against fabrication errors. Functional simulations were performed to validate the proposed scheme by configuring five distinct functionalities of varying complexity, including a Mach-Zehnder interferometer (MZI), a MZI lattice filter, a ring resonator (ORR), a double ORR ring-loaded MZI, and a triple ORR coupled resonator waveguide filter. The results demonstrate that our non-volatile scheme achieves performance equivalent to conventional PPICs. Robustness analysis was also conducted, and the results demonstrated that our scheme exhibits strong robustness against various fabrication errors. Furthermore, we explored the trade-off between the hardware design complexity of such non-volatile scheme and its performance. This study establishes a viable pathway to a new generation of power-connection-free PPICs, providing a practical and scalable solution for future photonic systems.
\end{abstract*}

\section{Introduction}
\begin{figure}[htbp]
\centering
\includegraphics[width=0.75\linewidth]{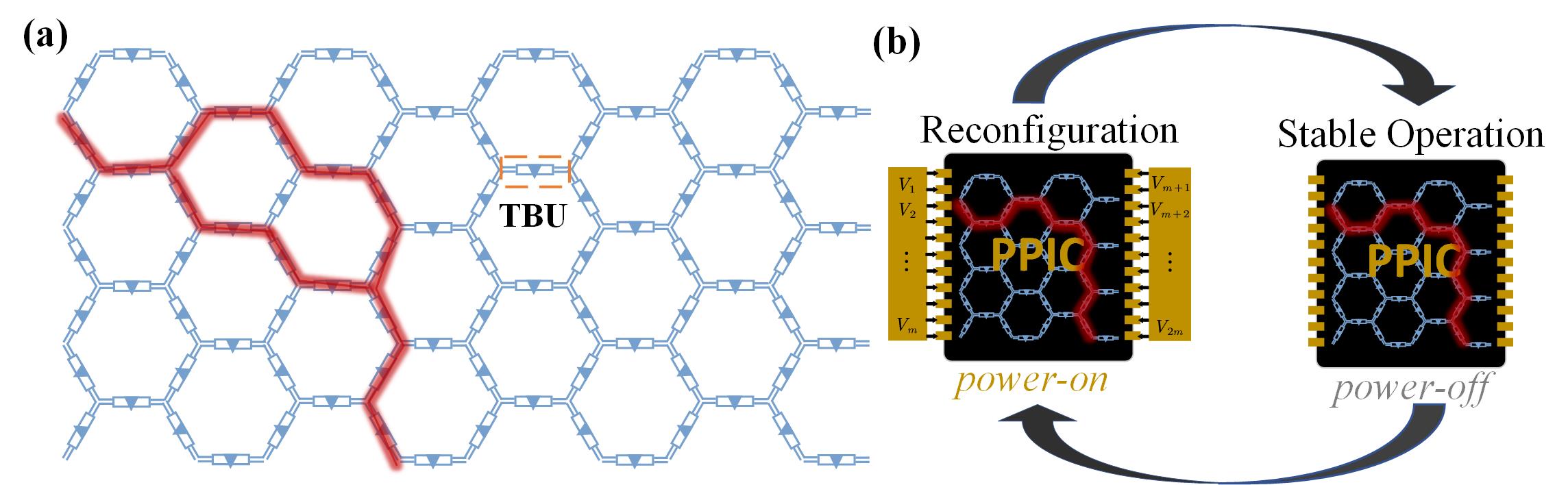}
\caption{(a) Programmable photonic circuit (PPIC) interconnected by many tunable basic units (TBUs). (b) Non-volatile actuators could enable a new generation of PPICs that can be (re)configured for different functionalities when powered and operate stably without any electrical connection.}
\label{fig:1}
\end{figure}

 Programmable photonic integrated circuits (PPICs) \cite{RN191,RN532,RN213,RN531} have emerged as a transformative platform in silicon photonics, offering a versatile alternative to application-specific photonic integrated circuits (ASPICs). Unlike ASPICs, whose structure varies with each user-defined functionality, PPICs employ a generic waveguide mesh interconnected by numerous tunable basic units (TBUs), as shown in Fig. \ref{fig:1}(a). These TBUs' coupling ratios and phase delays can be electrically controlled \cite{RN531}, thus allow diverse functionalities to be reconfigured on the same generic hardware by programming the electrical actuation, hence eliminates the need for multiple design and fabrication iterations for each functionality. Despite this revolutionary potential, PPIC technology remains immature, facing significant barriers to practical adoption.

One of the main challenges is scaling, as such generic PPICs by its definiation exhibit greater structural complexity then their ASPIC counterparts and enhanced functionality demands larger-scale circuits. Critically, the number of TBUs scales rapidly with circuit size, easily reaching hundreds or thousands \cite{RN788}. With so many electronically controled TBUs, power consumption becomes a critial issue, as thermo-optic actuators typically employed in current PPICs consume multiple mW of power per device\cite{RN788,RN790}.Therefore, photonic microelectromechanical systems (MEMS) actuators, as an power-effective tuning mechanism alternative, have garnered attention, and have been explored for PPICs implementations \cite{RN788,RN790,RN331,RN373,RN786}. However, such photonic MEMS actuators still require active biasing to maintain their states, limiting their usage to active operation.

Therefore, in this work, we propose a new PPIC architecture utilizing MEMS actuators with mechanical latching. Once a MEMS acutator is latched to a specific position, it maintains that state stably without requiring sustained electrical power \cite{RN331,RN792,RN787,RN785,RN375}. This enables a new generation of non-volatile PPIC that can operate as follows -- as conceptually illustrated in Fig. \ref{fig:1}(b) -- desired functionalities are (re)configured when powered, after which the system can operate persistently and stably without any electrical connection. This scheme not only helps reduce power consumption, but also enables a new, non-volatile passive operational paradigm, and elimiates the need for complex control system typically required to maintain stable states. 

The practical implementation of such non-volatile PPICs necessitates the redesign of both hardware and configuration algorithms. Therefore, in this work, we propose a practical system-level scheme including both (1) the non-volatile hardware design; and (2) an accompanying configuration algorithm allowing automatic defect-tolerant configuation of user-defined functionalities. The latching mechenism hardware design grant non-volatile operation, but constrained actuators to several specific latching positions, restricting TBUs to a limited set of discrete operationl states. The proposed configuration algorithm is desgined to compensate algorithmically for the inherent lack of continuous tunability to ensure performance. It is also capable of compensating for manufacturing variations to ensure reliable operation on imperfect fabricated chips.

To verify the effecitveness of the proposed scheme. We first perform functional simulations by configuring five distinct functionalities of varying complexity. This allows a direct comparison of the performance between the proposed non-volatile PPICs and conventional PPICs. Secondly, we conducted robustness analysis considering varies of fabrication errors to evaluate the error tolerance of the proposed scheme. At last, we explored the trade-off between the hardware design complexity of the non-volatile scheme and its performance.

\section{Principle}

\subsection{Hardware design of TBU using MEMS with mechanical latching}

\begin{figure}[htbp]
\centering
\includegraphics[width=0.75\linewidth]{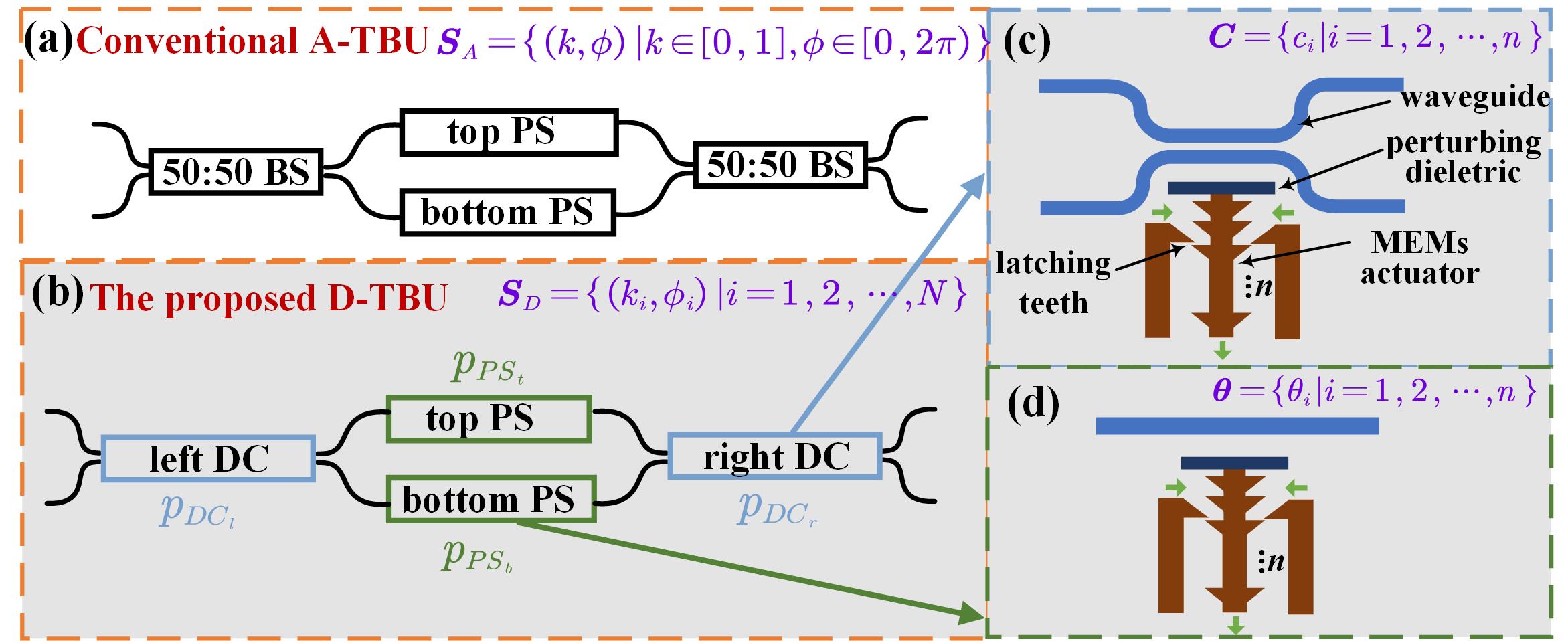}
\caption{(a) Conventional continuous tuning TBU allows achieving any arbitrary coupling ratio $k\in \left[ 0,1 \right]$ and phase delay $\phi \in \left[ 0,2\pi \right)$. (b) The proposed TBU hardware architecture comprising (c) tunable directional couplers (DC) and (d) phase shifters (PS) implemented with MEMS actuators with mechanical latching, each featuring $n$ latching positions, allowing TBU as a whole achieving $N$ different operational states $\boldsymbol{S}_D=\left\{ \left( k_i,\phi _i \right) \left| i=1,2,\cdots ,N \right. \right\} $.}
\label{fig:2}
\end{figure}

As shown in Fig. \ref{fig:2}(a), conventional TBUs typically employ a Mach-Zehnder interferometer (MZI) structures, incorporating two 50:50 beam splitters and continuously tuning phase shifters on both arms. The operational state of a TBU is generally characterized by its achieved coupling ratio $k$ and phase delay $\phi$. With such continuous tuning mechanism, such TBU are endowed with analog control to achieve any arbitrary coupling ratio between 0 and 1 and phase delay between 0 and 2$\pi$, in other words, granting TBU with full tuning freedom. Therefore, its set of possible operational states can be denoted as $\boldsymbol{S}_A=\left\{ \left( k,\phi \right) \left| k\in \left[ 0,1 \right] ,\phi \right. \in \left[ 0,2\pi \right) \right\} $.

In this work, we redesign the TBU structure by replacing the continuous tuning actuators with discrete tuning actuators implemented with MEMS with mechanical latching. As shown in Fig. \ref{fig:2}(b), the proposed D-TBUs employ the structure of a MZI consisting of two MEMS actuated tunable directional couplers (DC) and phase shifters attached to both arms. As shown in Fig. \ref{fig:2}(c)(d), such tunable PS/DC work by using a MEMS actuator to drive a movable perturbing dielectric towards or away from the optical waveguide\cite{RN774} \cite{RN331}, causing a change in the field profile. Specifically:

\begin{itemize}
\item for DC (Fig. \ref{fig:2}(c)), it creates a propagation constant mismatch between the adjacent waveguides, consequently altering the coupling ratio.
\item for PS (Fig. \ref{fig:2}(d)), this induces a phase shift in the propagating mode;
\end{itemize}

And the MEMS actuators utilizing mechanical latching \cite{RN331} to enable non-volatile operation by fixing the actuator at specific positions. As shown in Fig. \ref{fig:2}(c)(d), with $n$ discrete latching positions, a single MEMS tunable DC would be able to achieve $n$ possible distinct coupling ratios ($\boldsymbol{C}=\left\{ c_i\left| i=1,2,\cdots ,n \right. \right\} $) and a single MEMS PS can achieve $n$ possible different phase shifts ($\boldsymbol{\theta }=\left\{ \theta _i\left| i=1,2,\cdots ,n \right. \right\} $). Therefore, for a complete D-TBU with its left DC, right DC, top PS, and bottom PS each with $n$ latching positions, the D-TBU as a whole can achieve a total of $N=n^4$ distinct possible operational states. These states are characterized by the overall coupling ratio $k$ and phase delay $\phi$ of the D-TBU, forming a discrete set of possible operational states denoted as $\boldsymbol{S}_D=\left\{ \left( k_i,\phi _i \right) \left| i=1,2,\cdots ,N \right. \right\} $.

Clearly, such hardware design grant non-volatile operation, but constraint TBUs to a limited set of discrete operational states, thereby depriving them of the continuous tunability inherent in conventional TBU. However, this continuous tunability is fundamentally excessive, as in practice, optical signals propagate through a cascade of multiple TBUs along the propagation path, rather than an isolated single TBU. Consequently, even though an individual TBU is limited to specific discrete states, multiple TBUs along the optical path can be cooperatively adjusted using an algorithm to find a suitable combination that collectively achieve the user-defined functionality. The companion automatic configuration algorithm designed for this purposed will be introduced in the next section.

Note that for simplicity of expression, we refer to such discrete tuning TBUs as "digital" TBUs (abbreviated as D-TBUs), and PPICs composed of such TBUs as D-PPICs. In contrast, conventional continuous tuning TBUs and their corresponding PPICs are termed "analog" TBUs (A-TBUs) and "analog" PPICs (A-PPICs), respectively.

\subsection{Automatic configuration algorithm}

Here we introduce a configuration algorithm that can compensate the lack of continuous tunability caused by the non-volatile hardware design, ensuring user-defined functionalities can be automatically configured. Furthermore, the algorithm is engineered to be robust against fabrication imperfections, ensuring functions operate robustly and stably on defective chips.

Here we will detail the algorithm workflow, discuss the primary challenges in configuring D-PPICs, and explain how our algorithm overcomes them.

\begin{figure}[htbp]
\centering
\includegraphics[width=0.8\linewidth]{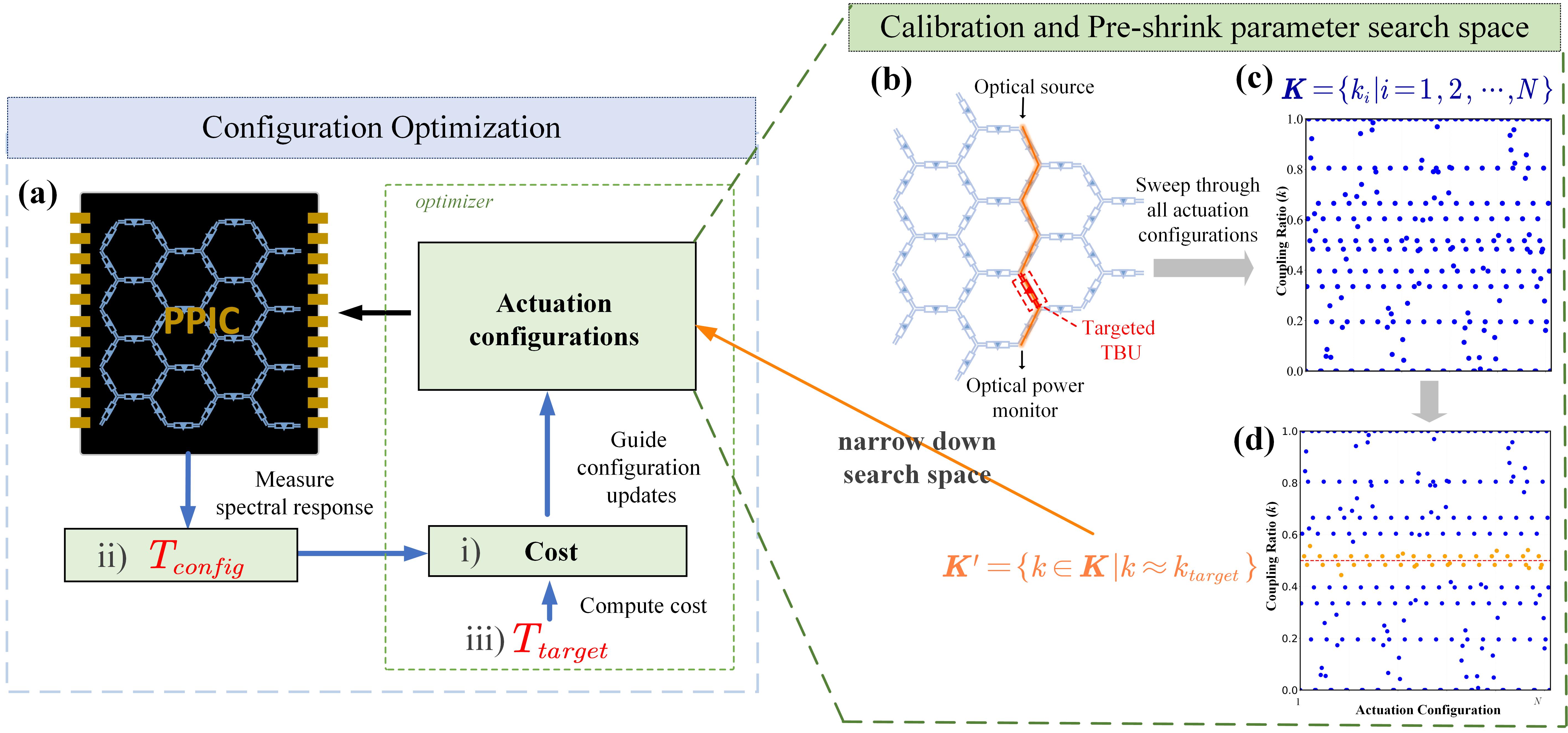}
\caption{Automatic configuration algorithm.}
\label{fig:3}
\end{figure}

As depicted in Fig. \ref{fig:3}(a), the configuration process essentially is to establish a feedback loop, a spectral mask ($T_{target}$) is defined for a specific user-defined functionality to serve as the configuration target for the D-PPIC to approximate. The spectral response of D-PPIC under current MEMS actuations configuration is monitored to serve as the feedback to guide the next step adjustment, thus iteratively refine the configuration until the spectral performance meet the target. Specifically, the actuation configurations of all relevant MEMS actuators are treated as parameters to be optimized, with the objective of minimizing the cost (see Fig. \ref{fig:3}(a) \text{i)}) -- defined as the difference between the spectral response achieved ($T_{config}$, see Fig. \ref{fig:3}(a) \text{ii)}) and the target spectral performance ($T_{target}$, Fig. \ref{fig:3}(a) \text{iii)})-- using a computational optimization algorithm. The spectral response is expected to converge to the target upon finding the optimal configuration.

While this algorithmic framework appears straightforward, directly applying such optimization is ineffective, as it faces three main issues that: (1) Configuration optimization of D-PPIC necessitates finding the global optimum rather than settling for a local optimum. This is because, the constrained discrete operational states of D-TBUs means that each TBU in the optical path can only achieve a few specific states. So, the optimization often must find the single, specific optimal solution that is both achievable for each TBU, and whose combination is collectively functionally adequate. This contrasts with their analog counterparts (A-TBUs), whose continuous modulation flexibility provides numerous possible combinations that can deliver the target functionality, making a local optimum often acceptable. This stringent requirement for a global optimum is particularly challenging given that PPIC configuration typically presents a non-convex optimization landscape, rendering algorithms highly susceptible to becoming trapped in local minima. (2) The discrete nature of D-TBUs introduces convergence challenges in optimization, as transitions between different operational states are abrupt. This results a highly non-smooth optimization landscape with sudden jumps, leading the optimization to struggles at progressing steadily and efficiently toward the optimum, ultimately impeding the convergence. (3) the configuration algorithm must be able to resist fabrication error to ensure practical applicability.

Therefore, to ensure the effectiveness of the configuration algorithm, we implemented the following key strategies. First, regarding the choice of optimization algorithm, we prioritize swarm intelligence optimization algorithms due to their strong global exploration capability and inherent robustness to discontinuities parameter landscapes, as opposed to gradient-based optimization methods which are susceptible to stuck in local optima and require continuity (erratic gradients would critically compromise convergence). Second, we introduce a step prior to the main optimization -- calibrating the actual fabricated chip and use the calibration results to narrow down the optimization parameter search space. This approach not only imbues the algorithm with robustness against fabrication errors but also significantly reduce the solution space the optimizer must explore, thereby lowering the overall optimization difficulty. The specific implementation workflow of this process will be introduced below.

\noindent(a) Calibration process

In the presence of fabrication error, the actual operational states $\boldsymbol{S}_D=\left\{ \left( k_i,\phi _i \right) \left| i=1,2,\cdots ,N \right. \right\}$ of a D-PPIC would deviate from the designed values and remain randomly unknown. However, in practice, although calibrating the actual phase delay ($\phi$) of TBU is difficult, calibrating the actual coupling ratio ($k$) is feasible \cite{RN550,RN540}. As illustrated in Fig. \ref{fig:3}(b)(c)(d), the calibration process can be summarized as follows: For a specific targeted TBU (highlighted in red), we employ an auto-routing algorithm \cite{RN544, RN577, RN290} to find a path that pass this TBU (orange path in Fig. \ref{fig:3}(b)). The transmission along this path is them maximized by setting all the TBUs along the path. Under this maximized condition, we sweep through all possible actuations on the constituent MEMS DC/PS in the targeted TBU while monitor the light power. These power measurements are then normalization and translated to TBU's coupling ratios, yield the set $\boldsymbol{K}=\left\{ k_i\left| i=1,2,\cdots ,N \right. \right\} $, which represents the actual operational coupling ratios. As shown in Fig. \ref{fig:3}(c), these blue dots represent the actual coupling ratios (y-axis) corresponding to all discrete actuation configurations (x-axis) of the D-TBU.

\noindent(b) Pre-shrink the parameter search space

Following the calibration, we can then narrow down the parameter search space based on the obtained calibration results. Specifically, a user-defined functionality can be translated into required coupling ratio and phase delay $\left( k_{traget},\phi _{target} \right)$ for each involved TBU. Obviously, for a D-TBU, it is barely possible to simultaneously reach both the exact target coupling ratio and phase delay. But with the calibration results providing the actual coupling ratio achieved under each actuation, we can screen and select only those actuations whose calibrated coupling ratio approximates the target ($\boldsymbol{K}' =\left\{ k\in \boldsymbol{K}\left| k\approx k_{target} \right. \right\} $). As illustrated in Fig. \ref{fig:3}(d), for a targeted coupling coefficient of 0.5, only the yellow points are retained after this screening process. This helps narrow down the parameter search space and the subsequent optimization only need to search within this pre-shrunk subset of actuation configurations. This ensures that the coupling ratio is inherently close to the target. The remaining deviations---including the phase delay error and residual coupling ratio inaccuracies---are then compensated by the following optimization process, which automatically coordinates within multiple TBUs on path.

In summary, the proposed the strategy of combining calibration with parameter space pre-shrinking not only ensures robustness against fabrication errors but also significantly reduces optimization complexity. Meanwhile, the final optimization process operates in a result-driven approach, thus enabling automatic and coordinated adjustments that guarantee the final functional performance.

\section{Result}

\subsection{Proof of Concept}

\begin{figure}[htbp]
\centering
\includegraphics[width=0.75\linewidth]{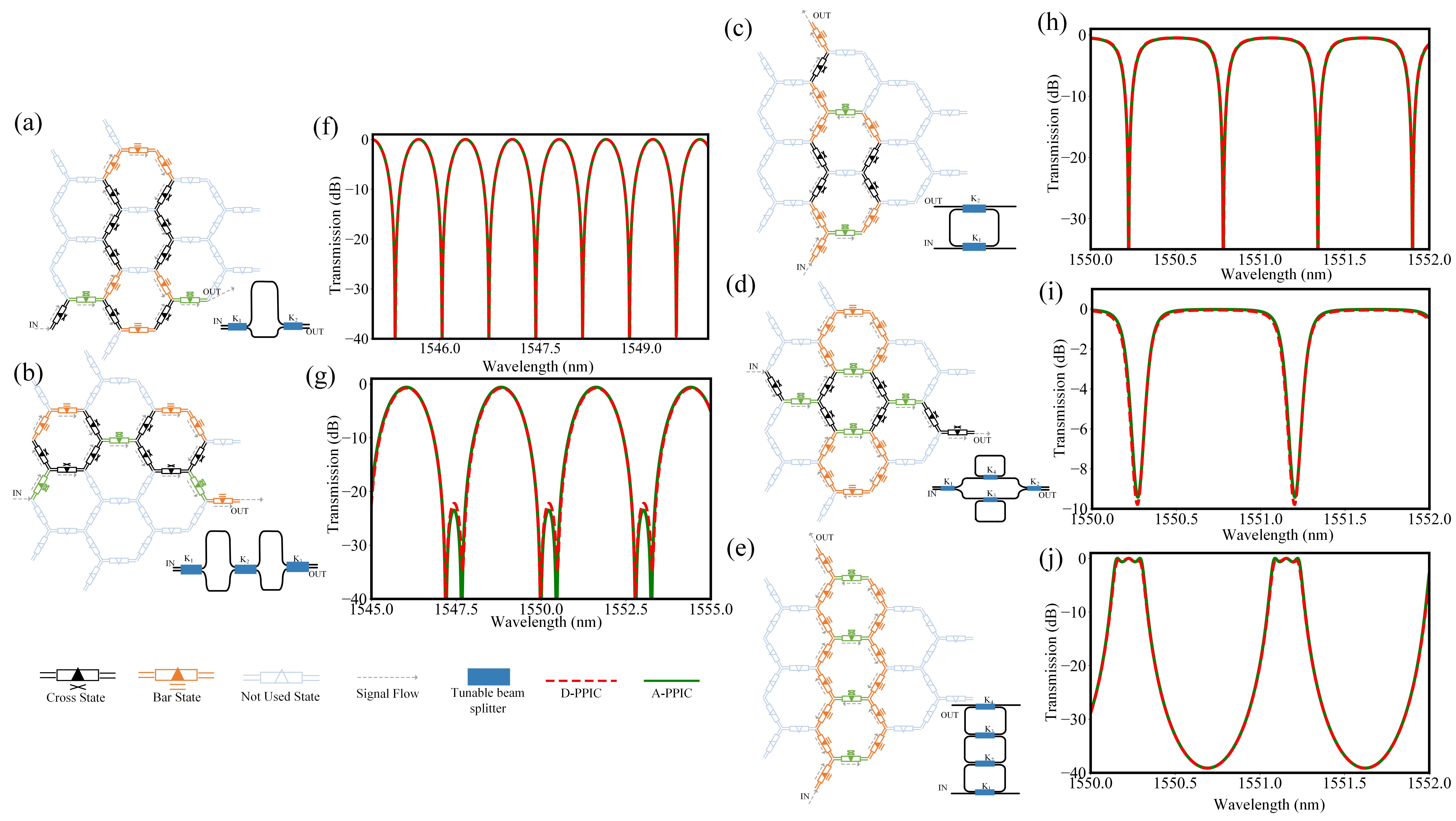}
\caption{Waveguide mesh connection diagrams, circuit layouts and configuration results for 5 different functionalities: (a) a Mach-Zehnder interferometer (MZI); (b) a MZI lattice filter; (c) a ring resonator (ORR); (d) a double ORR ring-loaded MZI; (e) a triple ORR coupled resonator waveguide filter. For each functionality, the spectral responses achieved (f)-(j) by conducting the configuration algorithm on a D-PPIC (red dashed) are compared with the spectral performance of a conventional A-PPIC (green solid).}
\label{fig:4}
\end{figure}

To evaluate the theoretical feasibility of the proposed scheme, we perform functional simulations by configuring multiple functionalities on both conventional A-PPICs and the proposed D-PPICs to compare their performance. As shown in Fig. \ref{fig:4}(a)-(e), five distinct functionalities of varying complexity were configured, including a Mach-Zehnder interferometer (MZI), a MZI lattice filter, a ring resonator (ORR), a double ORR ring-loaded MZI, and a triple ORR coupled resonator waveguide filter. Their configuration results on a D-PPIC ($n=4$ latching positions per MEMS actuator) and a conventional A-PPIC are compared in Fig. \ref{fig:4}(f)-(j), where green solid curve represents the spectral response of a conventional A-PPIC, and the red dashed curve depicts that achieved on a D-PPIC. It is evident that the two responses nearly overlap, indicating that the proposed D-PPIC working in conjunction with the automatic configuration algorithm, can deliver performance virtually identical to that of a conventional A-PPIC while achieving non-volatile operation.

\subsection{Robustness Analysis}

\begin{figure}[!htbp]
\centering
\includegraphics[width=0.76\linewidth]{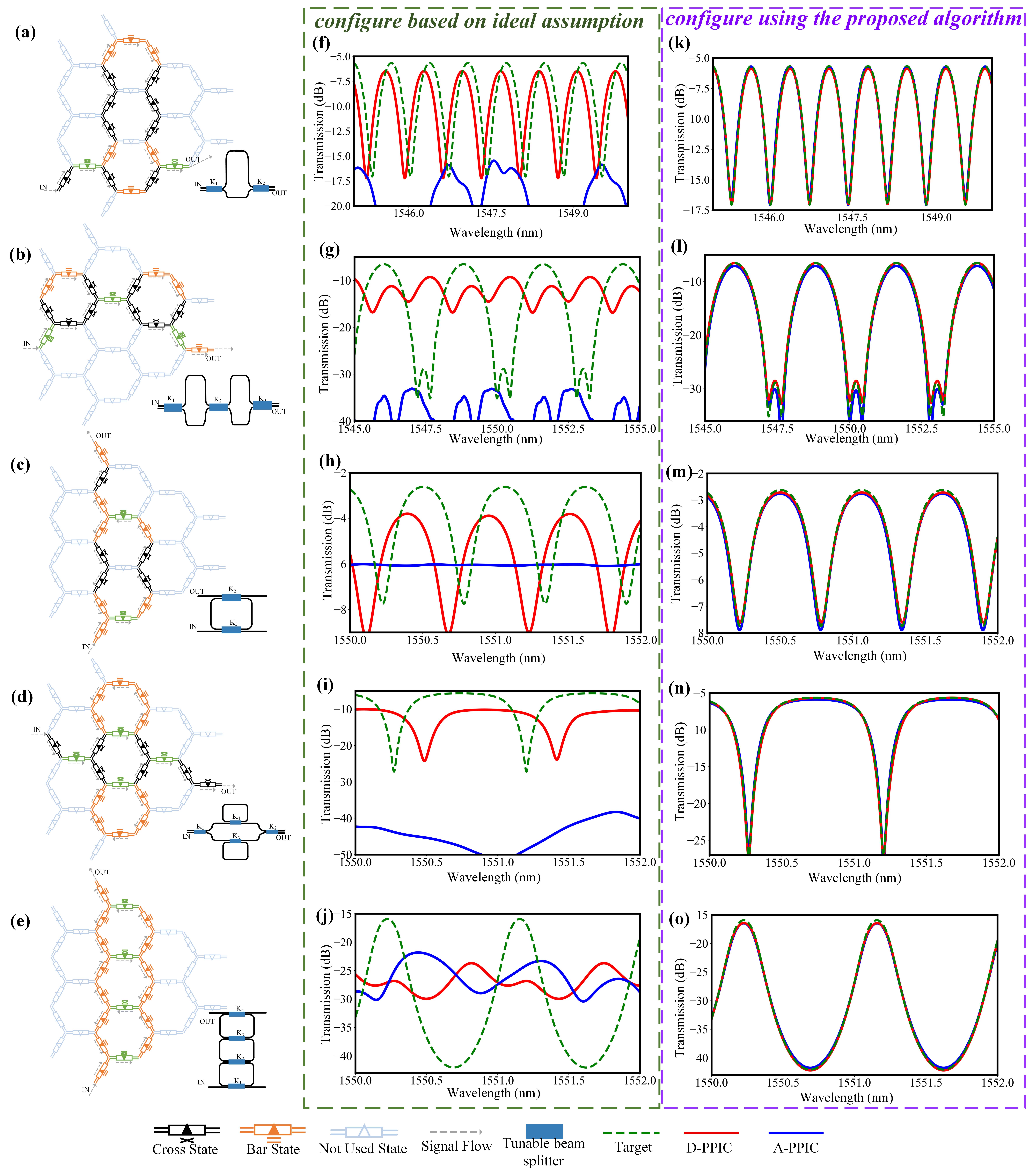}
\caption{Waveguide mesh connection diagrams, circuit layouts for five different functionalities: (a) a Mach-Zehnder interferometer (MZI); (b) a MZI lattice filter; (c) a ring resonator (ORR); (d) a double ORR ring-loaded MZI; (e) a triple ORR coupled resonator waveguide filter. Robustness analysis: under the realistic condition considering fabrication error: For each functionality, the spectral responses are compared between the target (green solid) and those achieved on an A-PPIC (blue dashed) and a D-PPIC (red dashed), configured respectively based on ideal error-free assumption (f)-(j) and using the proposed configuration algorithm (k)-(o). }
\label{fig:5}
\end{figure}

The practical fabrication of PPIC chips inevitably introduce deviations in key parameters due to manufacturing imperfections. To evaluate the robustness against fabrication errors of the proposed scheme, we implemented the previously mentioned five functionalities (Fig. \ref{fig:5}(a)-(e)) both on D-PPICs and A-PPICs under real-world conditions with fabrication errors, comparing their performance to assess relative robustness. The specific parameters settings modeling these fabrication errors are summarized in Table \ref{tab:1} in Appendix.

As shown in Fig. \ref{fig:5}(f)-(j), the green dashed curves represent the target spectral masks. The fabrication-induced errors would cause the actual operational states set $\boldsymbol{S}$ of the manufactured TBUs to deviate significantly from the designed one, which would lead the PPIC to malfunction if we still try to configure it based on an ideal error-free assumption, as shown in Fig. \ref{fig:5}(f)-(j), the configuration results we obtain on both D-PPICs (red solid curve) and conventional A-PPICs (blue solid curve) deviate completely from the target functionalities. However, the calibration incorporated in our proposed automatic configuration algorithm can effectively counteracts these fabrication errors, and the final result-driven optimization process can further mitigate such error effects. Therefore, as shown in Fig. \ref{fig:5}(k)-(o), when the proposed error-resilient automatic configuration algorithm is applied, the configuration results obtained on D-PPICs (red solid) and A-PPICs (blue solid) both closely match the target (green dashed), successfully achieving the desired functionality. To ensure comparability, the configuration algorithm we employed for A-PPICs closely mirrors the one used for D-PPICs. In D-PPIC case, we shrink the parameter search space by first ensuring the coupling ratios meet the target and only searching the subset actuations that satisfy coupling ratio. Similarly, for A-PPICs, we apply a similar search space reduction method by employing common-mode tuning of phase shifters on both arms of the A-TBU (Fig. \ref{fig:2}(a)) to maintain a constant coupling ratio that achieves the target. All other parts of the algorithm remain the same. Notably, in some of the more complex functions, the D-PPIC even slightly outperforms the A-PPIC, which can be attributed to the absence of thermal crosstalk and thermo-optic efficiency variations that inherently affect conventional A-PPICs.

\subsection*{Appendix}
Table \ref{tab:1} summarizes the modeled performance parameter deviations arising from different fabrication errors, along with their corresponding probabilistic representations.

\begin{table}[H]
    \centering
    \caption{Probabilistic models that different error parameters follow and their corresponding model}
    \label{tab:1}
    \footnotesize  
    \renewcommand{\arraystretch}{1.8}  
    \setlength{\tabcolsep}{4pt}        
    \setlength{\extrarowheight}{3pt}   
    \begin{tabular}{|>{\centering\arraybackslash}m{2.0cm}|>{\centering\arraybackslash}m{6.5cm}|>{\centering\arraybackslash}m{5.0cm}|}
    \hline
    \makecell[c]{\textbf{Fabrication}\\\textbf{errors}} & \makecell[c]{\textbf{Error parameters models}} & \makecell[c]{\textbf{Probabilistic models}} \\
    \hline
    \multirow{2}{2.0cm}{\\[1.5em]\centering Waveguide Dimension Errors and MEMS Latching Position Deviations}
        & \makecell[c]{$\mathbf{C} = \left\{ c_i \left| i=1,2,3,4 \right. \right\}$ \\[0.3em]
          \parbox[t]{6.2cm}{\justifying The actual operational coupling ratios of tunable directional couplers in D-PPICs deviate from their designed values due to: (1) MEMS actuator latching positions deviations, and (2) phase-matching inaccuracies in the directional couplers caused by waveguide dimensional variations.} \\[6em]} 
        & \makecell[c]{$c_i\sim \hat{c}_i\cdot \left( 1+N\left( 0,0.1^2 \right) \right)^a ,$ \\[0.3em]
          \parbox[c]{4.7cm}{\justifying  \noindent where $\hat{c}_i$ is the designed $i$-th possible coupling ratio of the tunable DC, and $c_i$ is the actual value.}} \\[0em]
    \cline{2-3}
        & \makecell[c]{$\boldsymbol{\theta }=\left\{ \theta _i\left| i=1,2,3,4 \right. \right\} $ \\[0.3em]
          \parbox[t]{6.2cm}{\justifying The actual operational phase shifts of phase shifters in D-PPICs deviate from their designed values, also due to MEMS actuator latching position deviations and waveguide dimension errors.} \\[1.5em]}
        & \makecell[c]{$\theta _i\sim \hat{\theta}_i\cdot \left( 1+N\left( 0,0.1^2 \right) \right)$, \\[0.3em]
          \parbox[c]{4.7cm}{\justifying \noindent where $\hat{\theta}_i$ is the designed $i$-th possible phase shift of the top/bottom PS, and $\theta _i$ is the actual value.}} \\[2em]
    \hline
    \multirow{3}{2.0cm}{\\[2em]\makecell[c]{\parbox[c]{1.8cm}{\centering Waveguide Dimension Errors}}}
        & \makecell[c]{$K_{BS}$ \\[0.3em]
          \parbox[t]{6.2cm}{\justifying The beam splitting ratio of beam splitters in A-TBU (Fig. \ref{fig:2} (a)) may deviate from the ideal 50:50 splitting \cite{RN552} due to waveguide dimension errors.}}
        & \makecell[c]{$K_{BS} \sim N\left(50\%,2.5\%^2 \right)$} \\[0.5em]
    \cline{2-3}
        & \makecell[c]{$\vartheta_e$ \\[0.3em]  
        \parbox[t]{6.2cm}{\justifying Fabricated waveguide dimension errors induce deviations in the effective refractive index, resulting in a passive phase offset ($\vartheta _e$) in the PSs (when not actuated).}}
        & \makecell[c]{$\vartheta _e\sim U\left[ 0,2\pi \right)^b $} \\[0.5em]
    \cline{2-3}
        & \makecell[c]{$n_g$ \\[0.3em]
          \parbox[t]{6.2cm}{\justifying Fabricated waveguide dimension errors induce deviations in group index \cite{RN551,RN554}.}}
        & \makecell[c]{$n_g\sim N\left( 4.3,0.01^2 \right) $} \\[0.5em]
    \hline
    \multirow{1}{2.0cm}{\\[0.0em]\makecell[c]{\parbox[c]{1.8cm}{\centering Thermo-Optic Actuator Imperfections}}}
        & \makecell[c]{$\eta_{TO}$ \\[0.3em]
          \parbox[t]{6.2cm}{\justifying The thermal-optic tuning efficiency of phase shifters in A-TBUs may deviate from design due to waveguide dimension errors, heater manufacturing error and material inhomogeneity \cite{RN794}.}}
        & \makecell[c]{$\eta _{TO}\sim \hat{\eta}_{TO}\cdot \left( 1+N\left( 0,0.05^2 \right) \right) $, \\[0.3em]
          \parbox[c]{4.7cm}{\justifying \noindent where $\hat{\eta}_{TO}$ is the designed tuning efficiency, and $\eta _{TO}$ is the actual value.}} \\[1.2em]
    \hline
    \multirow{1}{2.0cm}{\\[2em]\makecell[c]{\parbox[c]{1.8cm}{\centering Thermal Crosstalk}}}
        & \makecell[c]{$CT$ \\[0.3em]
          \parbox[t]{6.2cm}{\justifying  Crosstalk in thermo-optic phase shifters occurs when heat spreads to neighboring waveguides, producing an undesired tuning effect. This parasitic crosstalk can be modelled by a constant ($CT$) that reflects the percentage of phase shift occurred in the non-targeted waveguide \cite{RN556,RN557,RN565}. We account for the 10\% crosstalk between the upper and lower phase shifters within individual TBUs \cite{RN565}. Furthermore, the actual crosstalk exhibits random fluctuations around this nominal value due to unit-to-unit variations in thermo-optic tuning efficiency caused by fabrication imperfections. These efficiency variations result in differential phase responses to identical thermal stimuli across different phase shifters \cite{RN412,RN446}.}}
        & \makecell[c]{$CT\sim 10\%\cdot \left( 1+N\left( 0, 0.1^2 \right) \right) $, \\[0.3em]
          \parbox[c]{4.7cm}{\justifying \noindent where $CT$ is the thermal crosstalk coefficient with nominal value of 10\% and random variations.}} \\[0em]
    \hline
    \multirow{1}{2.0cm}{\\[1em]\makecell[c]{\parbox[c]{1.8cm}{\centering Optical Loss}}}
        & \makecell[c]{$IL$ \\[0.3em]
        \parbox[t]{6.2cm}{\justifying  Each TBU exhibits inherent insertion loss due to sidewall roughness scattering, material absorption and bend loss in fabricated waveguides \cite{RN533}.}}
        & \makecell[c]{$IL\sim U\left[ 0.5,0.7 \right] $, \\[0.3em]
        \parbox[c]{4.7cm}{\justifying \noindent where $IL$ is the insertion loss in dB .}} \\[1em]
    \hline  
    \end{tabular}
$^a$ $N\left( \mu ,\sigma ^2 \right) $ indicates a normal distribution with mean $\mu $ and standard deviation $\sigma $.\\
$^b$ $U\left[ a,b \right) $ indicates a uniform distribution within $\left[ a,b \right) $.
\end{table}

\subsection{Hardware Complexity and Performance Trade-off}

Clearly, the more complex the hardware design of the D-TBU -- that is, the greater the number of discrete operational states it can achieve -- the closer it approximates continuous tuning, thereby reducing the burden on the compensation algorithm in improving the overall performance. Therefore, we aim to explore the trade-off between the hardware design complexity of D-PPIC and its performance.

\begin{figure}[!htbp]
\centering
\includegraphics[width=0.70\linewidth]{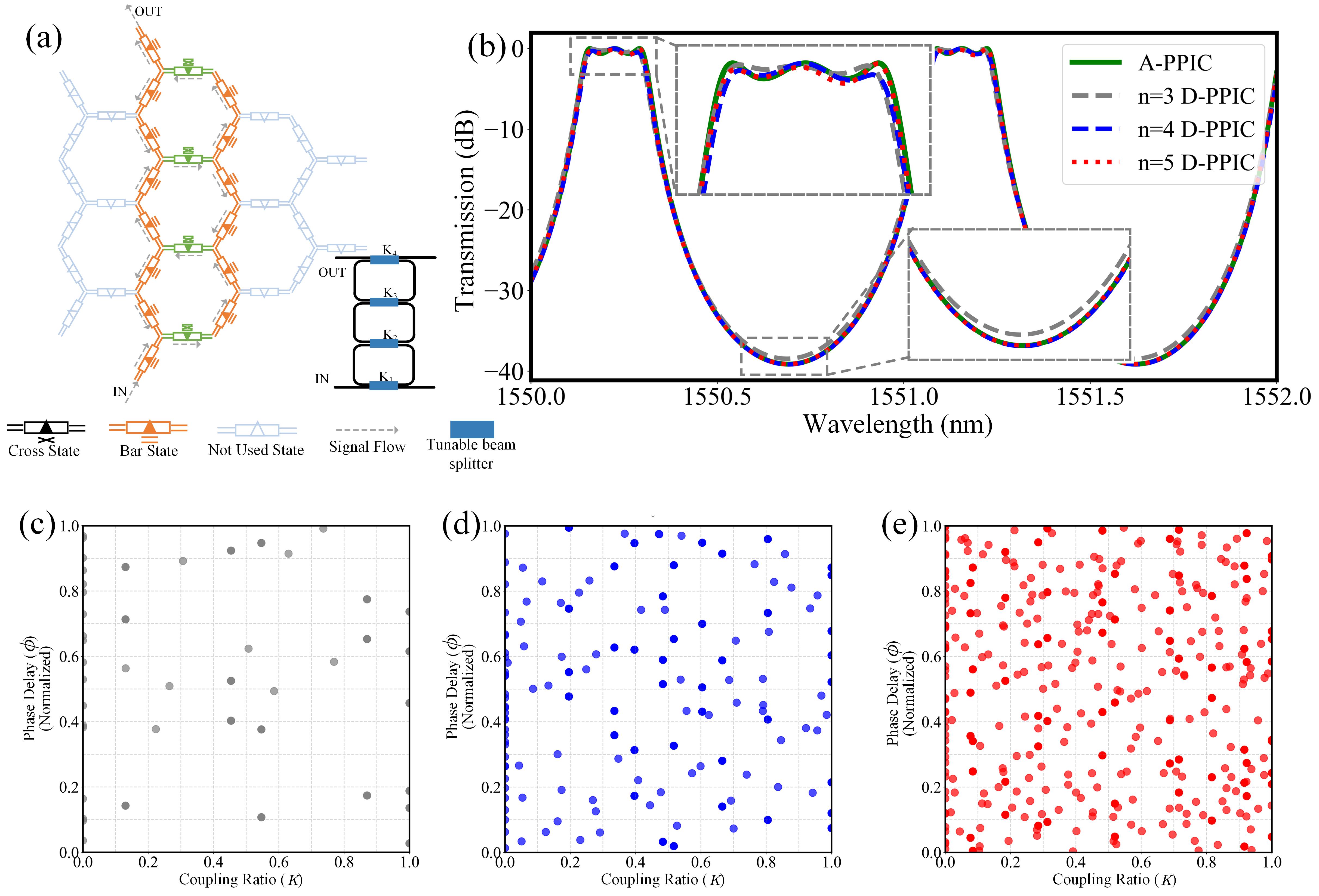}
\caption{(a) Waveguide mesh connection diagrams, circuit layouts of a triple ORR coupled resonator waveguide filter; (b) Configuration results of an A-PPIC, and D-PPICs with $n=3,4,5$, respectively. Distribution of operational states $\boldsymbol{S}_D$ in the $\left( k,\phi \right)$ plane of D-TBUs with (c) $n=3$; (d) $n=4$; (e) $n=5$;}
\label{fig:6}
\end{figure}

For a D-TBU where each constituent MEMS actuator possesses $n$ latching positions, the D-TBU as a whole would have $N=n^4$ discrete operational states $\boldsymbol{S}_D=\left\{ \left( k_i,\phi _i \right) \left| i=1,2,\cdots ,N \right. \right\} $. To determine the minimum required number of latching positions $n$ for satisfactory D-PPIC performance, we configured a complex triple ORR coupled resonator waveguide filter (Fig. \ref{fig:6}(a)) on D-PPICs with different $n$ and compared their performance. The results are presented in Fig. \ref{fig:6}(b), where the green solid curve represents the spectral performance of a conventional A-PPIC, and the gray dashed, blue dashed, and red dotted curves represent the configuration results of D-PPICs with $n=3,4$, and 5 respectively. As we can see, 4 latching positions are sufficient for D-PPIC to achieve nearly identical performance to A-PPIC, while 3 latching positions are already very close.

Notably, for any fixed number of latching positions ($n=3,4$, and 5), the TBU is carefully designed to maximizing the effective tuning freedom within the achievable range to best approximating continuous tuning. To explain in more detail, when visualize the $N$($N=n^4$) possible operational states by plotting each state coupling ratio $k$ on the x-axis and phase delay $\phi$ (normalized 0 to 1) on the y-axis (e.g., Fig. \ref{fig:6}(d) for $n=4$), a conventional A-TBU with full tuning freedom would continuously cover the entire plane. In contrast, the $N$ operational states of the D-TBU correspond to $N$ discrete points in this plane. As discussed earlier, we aim to maximize the effective tuning freedom -- that is, to maximize the $N$ operational states' coverage of the full tuning range -- rather than having them clustered in regions where multiple operational states yield similar coupling ratio and phase delay values. Graphically, this means we prefer the $N$ points to be uniformly distributed across the $\left( k,\phi \right)$ plane, rather than concentrated in specific areas. Thus, during D-TBU design, we perform such an optimization -- optimizing the latching positions of the constituent MEMS directional couplers and phase shifters within the D-TBU until the resulting operational states $\boldsymbol{S}_D=\left\{ \left( k_i,\phi _i \right) \left| i=1,2,\cdots ,N \right. \right\} $ achieve the most uniform distribution in the $\left( k,\phi \right)$. We applied this optimization process to D-TBUs of varying hardware complexity ($n=3,4, \text{and } 5$), and the resulting optimal states distribution for each case are shown in Fig. \ref{fig:6}(c)(d) and (e).

In summary, 4 latching positions per MEMS actuator is sufficient to guarantee performance comparable to conventional A-PPICs. Users can also choose the appropriate hardware complexity based on the desired performance level.

\section{Conclusion}

This study proposes a novel system-level design for non-volatile PPICs, utilizing MEMS actuators with mechanical latching, enabling power-connection-free operation while maintaining high performance. By replacing conventional continuous tuning actuators with discrete, latched MEMS actuators, and incorporating an automatic configuration algorithm, we successfully eliminate the need for sustained electrical connection without compromising performance.

By executing the proposed scheme to configure five different functionalities of varying complexity, we demonstrated that it can theoretically achieve performance comparable to conventional A-PPICs. Furthermore, robustness analysis was conducted to evaluate the scheme tolerance to fabrication errors, validating that the propose D-PPIC scheme exhibit robustness on par with A-PPICs. We also investigated the trade-off between hardware complexity and performance, concluding that merely 4 latching positions per MEMS actuator are sufficient. This work not only contributes to the ongoing development of scalable, low power consumption photonic systems but also provides a practical pathway for future PPICs with novel operational paradigm that allow non-volatile, power-connection-free operation.

\begin{backmatter}
\bmsection{Funding}
This work is partly supported by NSFC program 62275029.
 
\bmsection{Disclosures}
The authors declare that there are no conflicts of interest.

\bmsection{Data Availability Statement}
Data underlying the results presented in this paper are not publicly available at this time but may be obtained from the authors upon reasonable request.

\end{backmatter}

\bibliography{reference}

\end{document}